\documentclass[superscriptaddress,preprint,12pt]{revtex4-1}

\usepackage{amsmath}    
\usepackage{graphicx}   
\usepackage{verbatim}   
\usepackage{color}      
\usepackage{subfigure}  
\usepackage{hyperref}   
\usepackage{amssymb}
\usepackage{amsthm}
\usepackage{amsfonts}
\usepackage{float} 
\usepackage{natbib}

\begin{document}

\title{Gassmann Theory Applies to Nanoporous Media}

\author{Gennady Y. Gor} \email[Corresponding author, e-mail: ]{gor@njit.edu} \homepage[\\ URL: ]
{http://porousmaterials.net}
\affiliation{Otto H. York Department of Chemical, Biological, and Pharmaceutical Engineering, New Jersey Institute of Technology, University Heights, Newark, NJ 07102, USA}
\author{Boris Gurevich}
\affiliation{Curtin University, Perth, Western Australia, Australia}
\affiliation{CSIRO, Perth, Western Australia, Australia.}


\begin{abstract} 
Recent progress in extraction of unconventional hydrocarbon resources has ignited the interest in the studies of nanoporous media. Since many thermodynamic and mechanical properties of nanoscale solids and fluids differ from the analogous bulk materials, it is not obvious whether wave propagation in nanoporous media can be described using the same framework as in macroporous media. Here we test the validity of Gassmann equation using two published sets of ultrasonic measurements for a model nanoporous medium, Vycor glass, saturated with two different fluids, argon and n-hexane. Predictions of the Gassmann theory depend on the bulk and shear moduli of the dry samples, which are known from ultrasonic measurements, and the bulk moduli of the solid and fluid constituents. The solid bulk modulus can be estimated from adsorption-induced deformation or from elastic effective medium theory. The fluid modulus can be calculated according to the Tait-Murnaghan equation at the solvation pressure in the pore. Substitution of these parameters into the Gassmann equation provides predictions consistent with measured data. Our findings set up a theoretical framework for investigation of fluid-saturated nanoporous media using ultrasonic elastic wave propagation. 

\vspace*{1cm}
\noindent \textbf{Keypoints:}\\
 \begin{itemize}
 \item Bulk modulus of solid fraction can be calculated from adsorption-induced deformation
 \item Compressibility of confined fluid can be obtained from Tait-Murnaghan equation at solvation pressure
 \item Gassmann equation with the resulting moduli applies to ultrasound propagation in nanoporous media
 \end{itemize}

\vspace*{2cm}
\noindent \textbf{Note:} this is a preprint, the final version is published in \textit{Geophys. Res. Lett.} and available at \url{http://dx.doi.org/10.1002/2017GL075321}
\end{abstract}

\maketitle

\section{Introduction}
\label{sec:intro}

The last decade has seen enormous progress in production of hydrocarbons from unconventional reservoirs such as hydrocarbon-bearing shales and coal seams. Yet the spread of the technology worldwide and further gains in efficiency require application of geophysical reservoir characterization methods, which so far have been relatively ineffective in estimating spatial distribution of relevant formation properties. Application of geophysical methods to unconventional reservoirs requires thorough understanding of physical properties of these formations.  While there have been significant advances in both experimental and theoretical rock physics of unconventional reservoirs  (e.g., recent review by \citet{Pervukhina:etal:2015}), some fundamental questions are still poorly understood. In particular, it is not known whether the Gassmann theory, which plays a central role in seismic characterization of conventional reservoirs \citep{Smith2003}, applies to nanoporous materials (Following the most recent IUPAC convention we call here ``nanoporous materials'', those materials that have the pore sizes below 100~nm \citet{Thommes2015}). Indeed, both shales and coal seams are known to have a nano-scale structure. Many mechanical and thermodynamic properties of nano-scale solids and fluids differ from the bulk properties of the same materials \citep{Nastasi2012, Huber2015}. Therefore, applicability of conventional rock physics theories such as the Gassmann theory \citep{Gassmann1951} to these materials is uncertain. As an amusing example, a debate at a
workshop on rock physics of unconventional reservoirs held by the Society of Exploration Geophysicists (Denver, Colorado, October 2014)  led to a vote of the attendees on whether they believe the Gassmann theory is applicable to shales; and the opinions of the attendees were evenly split. 

Testing the Gassmann theory experimentally on unconventional reservoir rocks is challenging for a number of reasons. First, direct replacement of fluids in such a complex system would trigger a number of physicochemical processes, some of which may be irreversible \citep{Rezaee:2015}. Second, some of the key minerals in shales (e.g. clays) only exist as nano-scale particles and their mechanical properties are a subject of long lasting debates \citep{Vanirio:Prasad:Nur:2003,Mondol:etal:2008}. Third, these minerals are strongly anisotropic \citep{Sayers:denBoer:2016}, and hence require the use of an anisotropic version of the Gassmann theory \citep{Brown:Korringa:1975}, which involves a greater number of parameters. Therefore, as a first step towards resolving this question, it is useful to study simplified model systems, such as nanoporous glass. This has been done by \citet{Page1995} and \citet{Schappert2014}, who measured longitudinal and shear ultrasonic velocities in nanoporous Vycor glass as a function of  pressure of n-hexane and argon vapors respectively. Both studies show that when the vapor pressure reaches a capillary condensation point, the longitudinal modulus of the sample increases sharply and this increase is in a broad agreement with the Gassmann theory. However, quantitative testing of the Gassmann theory on Vycor glass requires the knowledge of the bulk modulus of the solid nonporous glass $K_{\rm{s}}$, which cannot be measured directly. 

In this paper we propose to estimate $K_{\rm{s}}$ from the so-called pore-load modulus obtained from measurements of the adsorption-induced deformation. Furthermore, comparison of this estimate with the estimates from the elastic effective medium theory gives a recipe for estimating $K_{\rm{s}}$ when adsorption-induced deformation data are not available. We then substitute the estimated values of $K_{\rm{s}}$ along with the longitudinal modulus of the dry porous glass and the fluid compressibility into the Gassmann equation to compute the saturated (undrained) longitudinal modulus of the porous glass filled with liquid argon or n-hexane. Finally we verify the validity of the Gassmann theory by comparing these predictions against the ultrasonic measurements of \citet{Page1995} and \citet{Schappert2014}.

\section{Gassmann Model}
\label{sec:Gassmann}

We consider a porous solid with porosity $\phi$, which can be saturated with a fluid. We introduce the following elastic properties to describe it: $M$ is the longitudinal modulus, $G$ is the shear modulus and $K = M - \frac{4}{3} G$ is the bulk modulus. All these moduli correspond to adiabatic conditions.

In the low-frequency limit, the bulk modulus of the fluid-saturated medium $K$ is related to its dry bulk modulus $K_{\rm{0}}$, solid modulus $K_{\rm{s}}$ and fluid modulus $K_{\rm{f}}$ by the 
\citet{Gassmann1951} equation (see also \citep{Berryman1999})
\begin{equation}
\label{Gassmann}
K = K_0 + \frac{\left( 1 - \frac{K_0}{K_{\rm{s}}}\right)^2}{\frac{\phi}{K_{\rm{f}}} + \frac{1-\phi }{K_{\rm{s}}} - \frac{K_0}{K_{\rm{s}}^2}},
\end{equation} 
while the shear modulus of the saturated sample $G$ equals to the that of the dry material $G_0$,
\begin{equation}
\label{GassmannG}
G=G_0.
\end{equation} 
The values without subscripts are the properties of the composite (saturated porous body), the values with the subscript ``$0$'' correspond to the dry (drained) porous body, the superscripts ``$\rm{s}$'' and ``$\rm{f}$'' denote the properties of the solid non-porous material and fluid, respectively. Since the experimentally measured quantity is the longitudinal modulus of the composite, it is convenient to rewrite Eq. \ref{Gassmann} as
\begin{equation}
\label{GassmannM}
M = M_0 + \frac{(K_{\rm{s}} - K_0)^2 K_{\rm{f}}}{\phi K_{\rm{s}}^2 + \left[ (1 - \phi )K_{\rm{s}} - K_0 \right] K_{\rm{f}}}.
\end{equation}

Computing the longitudinal modulus of the saturated samples using Eq. \ref{GassmannM} requires the knowledge of the porosity of the material $\phi$, moduli of the dry sample $M_0$ and $G_0$ (which can be readily calculated from the travel times of the longitudinal and transversal waves), fluid modulus $K_{\rm{f}}$ (usually known at the current pressure and temperature \citep{Batzle:Wang:1992, Smith2003}), and measured solid modulus $K_{\rm{s}}$. However, estimation of the solid modulus $K_{\rm{s}}$ (often called mineral or matrix modulus) is often a non-trivial problem. For multi-mineral rocks, this requires the use of an approximate elastic effective medium theory (EMT) \citep{Smith2003}. The problem is even more challenging when the nature of the material precludes the existence of macroscopic non-porous samples that could be measured directly. This is the situation with shales or mudrocks as clay minerals usually exist as nano-scale particles \citep{Vanirio:Prasad:Nur:2003,Mondol:etal:2008}. Similarly, the technology for producing the Vycor glass makes it porous, and use of the conventional (nonporous) glass properties to describe Vycor's solid fraction is problematic, as these properties strongly depend on the glass's thermal history \citep{Scherer1986}. However there exists a potential to estimate this modulus from adsorption-induced deformation experiments \citep{Gor2017review}. This approach is described in the next section.

\section{Methods}
\label{sec:Methods}

\subsection{Adsorption Experiments on Vycor Glass}
\label{sec:adsorption}

\begin{figure}[ht!]
\centerline{\includegraphics[height=1.5in]{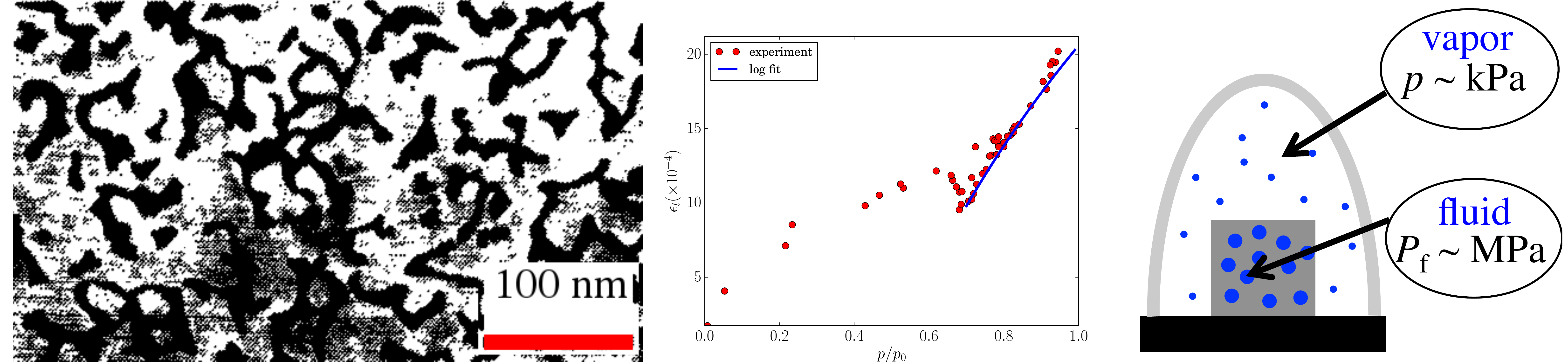}}
\caption{(Left) Microstructure of Vycor 7930 porous glass, reproduced with permission from \citep{Levitz1991} (copyright AIP Publishing LLC). (Center) Adsorption-induced linear strain of Vycor glass sample measured during water vapor adsorption at 18.75$^{\circ}$C \citep{Amberg1952} plotted as a function of relative vapor pressure $p/p_0$, where $p_0 = 2.16$~kPa is the saturated vapor pressure. The solid line shows the fitting to a $\log$ function. (Right) Schematic of the gas adsorption experiment, demonstrating the difference between the two key quantities: vapor pressure $p$ and fluid pressure in the pores $P_{\rm{f}}$.}
\label{fig:porous}
\end{figure}

The analysis of the applicability of the Gassmann theory will be based on ultrasonic measurements carried out during adsorption experiments on a model nanoporous system, Vycor glass. Vycor is a high-silica borosilicate glass, with porosity around 30\%, which has pores with disordered worm-like structure, 6-8 nanometers in diameter. The left panel of Figure \ref{fig:porous} shows the structure of Vycor based on a combination of experimental characterization techniques \citep{Levitz1991}. Vycor has high porosity and narrow pore size distribution; these features made it widely used as a medium for studying confinement effects for various solid and fluid phases \citep{Schappert2015EPL,  Schappert2016JPCC, Gruener2016, Uskov2016, Soprunyuk2016, Filippov2016, Egorov2017}. Note also that combination of various experimental methods used to calculate the porosity of Vycor suggests that Vycor does not have isolated (occluded) porosity \citep{Levitz1991, Page1995}.

\begin{figure}[ht!]
\centerline{\includegraphics[height=2in]{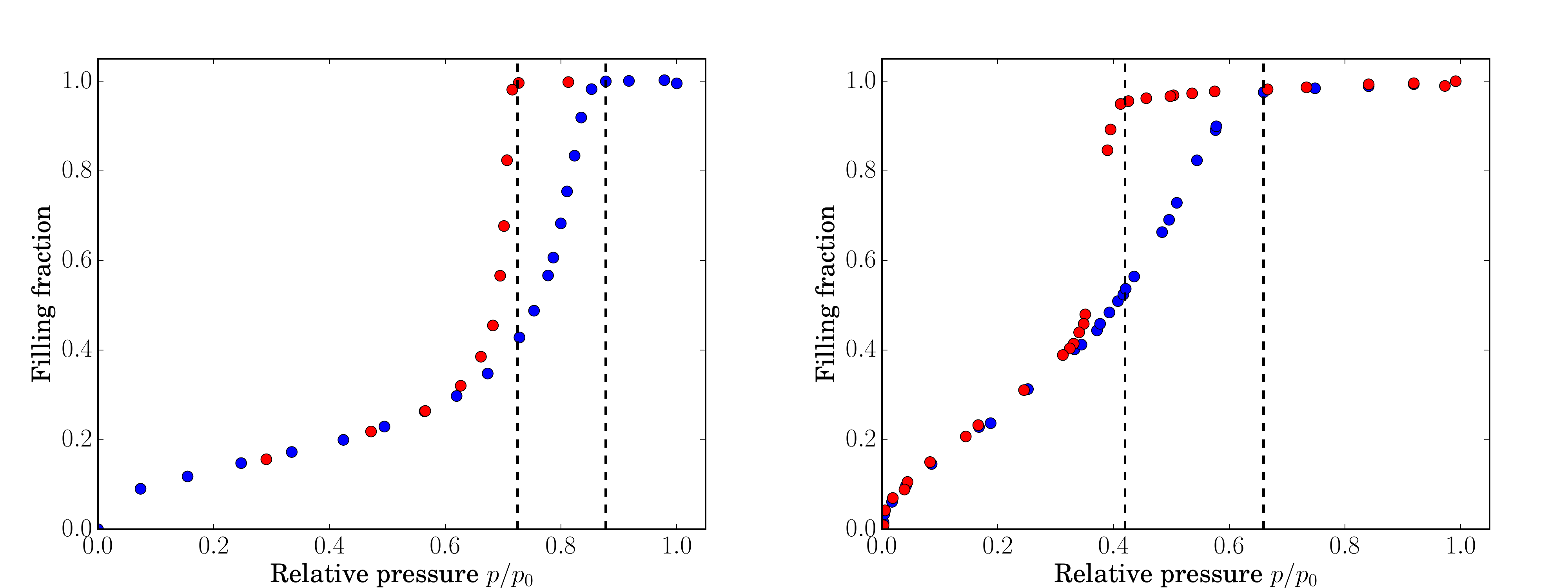}}
\caption{Experimental adsorption isotherms of different vapors on Vycor glass: filling fraction of the total pore volume as a function of relative vapor pressure $p/p_0$ at the constant temperature $T$.  Left panel shows the data for argon adsorption at $T=80$~K from \citet{Schappert2014}. The right panel shows the data for n-hexane from \citet{Page1995}. The vertical dashed lines mark the points of capillary condensation ($p_c$) and evaporation ($p_e$).}
\label{fig:isotherms}
\end{figure}

Starting from the 1980s, several groups used Vycor for ultrasonic investigation of nanoconfined fluids and solids \citep{Murphy1982, Beamish1983, Warner1988, Page1995, Charnaya2001, Schappert2014}. We discuss here the experiments in which the pores in Vycor are filled with fluids. In these experiments a dry sample is gradually saturated by means of vapor adsorption at a constant temperature (schematic of such experiments is shown in the right panel of Figure \ref{fig:porous}). Vapor adsorption in a mesoporous material takes place as the following: at low vapor pressures $p$, the fluid adsorbs as a liquid-like film on the pore walls surface and the thickness of this film gradually grows with the increase of the vapor pressure. After a certain pressure value, $p_c$, which is lower than the saturated vapor pressure $p_0$, capillary condensation takes place and the entire pore gets filled with the condensate. Here we consider only the part of the experimental data beyond the capillary condensation, where the system probed by ultrasound consists of only two phases: solid matrix and nanoconfined liquid-like fluid in the pores. If in an adsorption experiment the vapor pressure is decreased from the saturated state, the pores will empty at a certain pressure $p_e$. Typically $p_e < p_c$ and the adsorption isotherms have a hysteresis loop as in Figure \ref{fig:isotherms}.  

For our analysis we use published data from \citep{Page1995} and \citep{Schappert2014}. The former is concerned with n-hexane adsorption at room temperature ($T = 295.75$~K) and the latter uses argon at $T=80$~K. The adsorption isotherms from these works are shown in Figure \ref{fig:isotherms}. The vertical dashed lines show the capillary condensation and evaporation points. 

\subsection{Solid Bulk Modulus of Nanoporous Glass}
\label{sec:Modulus}

As mentioned earlier, the Gassmann equation, Eq.~\ref{Gassmann}, requires the knowledge of the solid modulus $K_{\rm{s}}$, which cannot be measured directly, but can be estimated using the following approach. When fluids saturate nanoporous solids they cause noticeable strains, typically of the order of $10^{-4}-10^{-3}$. This phenomenon is known as adsorption-induced deformation and is well documented in the literature (see the review \citep{Gor2017review} and references therein). Adsorption-induced deformation is notorious for causing coal swelling and shrinkage in enhanced coal-bed methane recovery \citep{Yang2011,Espinoza2013,Zhang2016}. 

The fluid pressure in the pores $P_{\rm{f}}$, which is the driving force for the strain, is called ``solvation pressure'' or ``adsorption stress''. Typically in sub-critical adsorption experiments, solvation pressure $P_{\rm{f}}$ exceeds the external (vapor) pressure $p$ by several orders of magnitude \citep{Gor2010} (see Figure \ref{fig:porous}). Thus the difference between the adsorption stress and solvation pressure is negligible. For mesoporous materials, i.e when pores are in the range of $2-50$~nm \citep{Thommes2015}, the solvation pressure in the pores filled with capillary condensate can be written in macroscopic terms as \citep{Gor2010,Gor2014}
\begin{equation}
\label{Psolvation}
P_{\rm{f}} = P_{\rm{sl}} + P_{\rm{L}}
\end{equation} 
where the first term is related to solid-fluid interactions and the second term is the Laplace pressure
\begin{equation}
\label{PLaplace}
P_{\rm{L}} = \frac{R_{\rm{g}} T}{V_{\rm{l}}} \log\left( \frac{p}{p_0}\right).
\end{equation} 
Here $R_{\rm{g}}$ is the gas constant, $T$ is the absolute temperature and $V_{\rm{l}}$ is the molar volume of the liquid phase.  Note that while the first term in Eq.~\ref{Psolvation} is compressive, the second term causes the tensile stresses when the system is in equilibrium with undersaturated vapor at $p < p_0$; at $p=p_0$ the Laplace pressure term vanishes.

For a given sample saturated with a fluid, the $P_{\rm{sl}} = \rm{const}$, so Eq.~\ref{PLaplace} gives the simple logarithmic dependence of linear strain of the porous sample $\epsilon_l$ on $p/p_0$, shown in Figure \ref{fig:porous} and observed for all mesoporous materials \citep{Gor2017review}. This dependence is often described using a special elastic modulus related to this process, so-called ``pore-load modulus'' $\mathcal{M}_{\rm{PL}}$ \citep{Prass2009,Gor2015modulus} as a proportionality constant in the linear relation between the solvation pressure $P_{\rm{f}}$ and measured $\epsilon_l$. $\mathcal{M}_{\rm{PL}}$ can be  related to elastic moduli using Biot's theory of poroelasticity. According to Biot's theory, the volumetric strain $\epsilon$ is a function of hydrostatic effective stress (compressive stresses are assumed positive) \citep{Nur1971, Rice1976, Berryman1993, Detournay1993}
\begin{equation}
\label{Pe}
\sigma_{\rm{e}}=P_{\rm{c}}-\alpha P_{\rm{f}}, 
\end{equation}
where $P_{\rm{c}}$\ is confining pressure, $P_{\rm{f}}$ is the fluid pressure in the pores, and $\alpha$ is the Biot-Willis effective stress coefficient. For a porous material made up of a single isotropic solid substance with bulk modulus $K_{\rm{s}}$, 
\begin{equation}
\label{alpha}
\alpha =1-K_{0}/K_{\rm{s}},  
\end{equation}
where $K_{0}$ is the drained bulk modulus 
\begin{equation}
\label{K}
\frac{1}{K_{0}}=-\left. \frac{\partial \epsilon}{\partial P_{\rm{c}}}\right\vert
_{P_{\rm{f}}=0}.  
\end{equation}
Similarly, the pore load modulus can be written as
\begin{equation}
\label{PLM}
\frac{1}{\mathcal{M}_{\rm{PL}}} = \left. \frac{\partial \epsilon_{l}}{\partial P_{\rm{L}}} \right\vert _{P_{\rm{c}}} 
= \left. \frac{\partial \epsilon_{l}}{\partial P_{\rm{f}}} \right\vert _{P_{\rm{c}}} 
= \frac{1}{3}\left. \frac{\partial \epsilon}{\partial P_{\rm{f}}}\right\vert _{P_{\rm{c}}},  
\end{equation}
where we took into account that the strains are small so that the volumetric strain $\epsilon = 3 \epsilon_{l}$. 

Since the confining pressure is the vapor pressure in the adsorption experiment $P_{\rm{c}} = p < 1$~atm, it can be assumed zero, therefore
\begin{equation}
\label{dedP}
\left. \frac{\partial \epsilon}{\partial P_{\rm{f}}}\right\vert _{P_{\rm{c}}} =
\frac{d\epsilon }{d\sigma_{\rm{e}}}\frac{\partial \sigma_{\rm{e}}}{\partial P_{\rm{f}}} =-\alpha \frac{d\epsilon }{d\sigma_{\rm{e}}}.
\end{equation}
Then, using Eqs \ref{Psolvation}, \ref{Pe}, \ref{dedP} and \ref{PLM}, Eq.~\ref{K} gives 
\begin{equation}
K_{0}^{-1}=-\left. \frac{\partial \epsilon}{\partial P_{\rm{c}}}\right\vert
_{P_{\rm{f}}}=-\frac{d\epsilon}{d\sigma_{\rm{e}}}=\frac{1}{\alpha }\left. \frac{\partial \epsilon}{\partial P_{\rm{f}}}\right\vert_{P_{\rm{c}}}= 3\alpha^{-1}\mathcal{M}_{\rm{PL}}^{-1},
\end{equation}
or
\begin{equation}
\label{eq_for_K}
K_{0}=\frac{\left( 1-K_{0}/K_{\rm{s}}\right) \mathcal{M}_{\rm{PL}}}{3},  
\end{equation}
so that 
\begin{equation}
\label{eq_for_K1}
\frac{3}{\mathcal{M}_{\rm{PL}}}+\frac{1}{K_{\rm{s}}}=\frac{1}{K_{0}}. 
\end{equation}
When $\mathcal{M}_{\rm{PL}}$ and $K_0$ are known from measurements, then Eq.~\ref{eq_for_K1} can be used to estimate $K_{\rm{s}}$. This equation was derived by \citet{Mackenzie1950} for spherical pores, but the above derivation shows that it is valid for any pore geometry as long as all the pores are inter-connected.

The advantage of estimating $K_{\rm{s}}$ in this way is that it is independent of pore geometry and does not require the knowledge of porosity or pore fluid properties. Alternatively $K_{\rm{s}}$ can be estimated from porosity $\phi$, bulk and shear moduli of the dry sample $K_0$ and $G_0$ using the EMT \citep{Kuster:Toksoz:1974,Berryman1980} and assuming that the pores are approximately cylindrical in shape. The details of this calculation are given in Supporting Information. It might also be possible to estimate the solid bulk modulus from forced confining and pore fluid pressure oscillation measurements \citep{Pimienta2017}. However for nanoporous media this may be problematic. Indeed, this would need to be done with a non-sorbing fluid, and sorption is an inherent feature of nanopores for almost any fluid.

\subsection{Bulk Modulus of the Fluid in the Pores}
\label{sec:Tait}

For macroporous media Eq \ref{GassmannM} can be used assuming the properties of solid and liquid constituents being the same as the bulk properties at the same temperature and pressure. In particular, the bulk modulus of the liquid in the pores $K_{\rm{f}}$ can be taken from the reference thermodynamic data for a given fluid. However, nanoconfinement is known to alter thermodynamic properties of fluids \citep{Huber2015}. In particular the fluid in the nanopores is at a pressure $P_{\rm{f}}$ different from the pressure $p$ of the equilibrium bulk phase (Figure SI.1 in the Supporting Information shows the plot of $P_{\rm{f}}$ as a function of $p$ for argon in model Vycor pores). Therefore, the effects of the nanoconfinement are manifested in the pressure dependence of these thermodynamic properties. The pressure dependence of the bulk modulus of a bulk liquid is described by a classical Tait-Murnaghan equation \citep{Murnaghan1944, Birch1952}
\begin{equation}
\label{TaitF}
K_{\rm{f}}(P_{\rm{f}}) \simeq K_{\rm{f}}(0) + K_{\rm{f}}^{\prime}P_{\rm{f}},
\end{equation}
where $K_{\rm{f}}^{\prime} = d K_{\rm{f}}/dP_{\rm{f}}$, which is constant for a given liquid in a wide range of pressures. Recent molecular simulations for simple fluids show that the Tait-Murnaghan equation remains valid when the fluids are confined in the nanopores \citep{Gor2014, Gor2015compr, Keshavarzi2016}. Moreover, the ``slope'' $K_{\rm{f}}^{\prime}$ in  Eq.~\ref{TaitF} for confined argon is the same as for bulk, \citep{Gor2016Tait}, and the latter can be readily calculated from reference thermodynamic data. Therefore, knowing the solvation pressure $P_{\rm{f}}$ from macroscopic or molecular theories of adsorption \citep{Gor2011}, one can calculate the modulus of the confined fluid in the pores from Eq.~\ref{TaitF}.

\section{Results}
\label{sec:Results}

To test our method we consider ultrasonic experiments carried out during adsorption of vapors on Vycor samples.

\subsection{Properties of Vycor Samples}
\subsubsection{Sample from \citet{Schappert2014}}

\citet{Schappert2014} used Vycor glass No. 7930. The same type of glass was also used by \citet{Amberg1952} to measure deformation induced by adsorption of water. These data are plotted in Figure \ref{fig:porous} and can be used to estimate the solid bulk modulus of Vycor. The $\log$-fitting for the system shown in Figure \ref{fig:porous} gives pore-load modulus $\mathcal{M}_{\rm{PL}} = 44.5$~GPa. The drained bulk modulus measured by \citet{Schappert2014} is $K_{0}=7.73$~GPa. Substitution of these values into Eq.~\ref{eq_for_K1} gives $K_{\rm{s}}=16.1$~GPa.

The obtained value of $K_{\rm{s}}$ for Vycor is noticeably smaller than the bulk modulus of $\alpha$-quartz, $K_{\rm{s}} = 37.41$~GPa \citep{Anderson1966}. It is close, however, to value $K_{\rm{s}} = 17.95$~GPa calculated by \citet{Scherer1986} from the dry moduli using elastic EMT (\citet{Scherer1986} reports the Young's modulus $E_{\rm{s}} = 37.7$~GPa and Poisson's ratio $\nu = 0.15$). According to \citet{Scherer1986}, the modulus of solid Vycor matrix is lower than for quartz due to the presence of large amount of hydroxyls. When the sample is heated to 550-800$^{\circ}$C, the glass gets dehydroxylated, and the modulus approaches the quartz values. 

In addition to the bulk modulus $K_0$, \citet{Schappert2014} report the shear modulus of the dry sample, $G_{0}=6.86$~GPa. Using these values of $K_{0}$ and $G_{0}$, and porosity $\phi =0.28$, and assuming that the pores are infinite circular cylinders, we can estimate the moduli based on the EMT, getting $K_{\rm{s}}=14.13$ GPa and $G_{\rm{s}}=13.97$ GPa. The details of this estimation are given in Supporting Information.

The $K_{\rm{s}}$ value obtained from the effective stress is independent of porosity, pore geometry and fluid compressibility. However, it is sensitive to $\mathcal{M}_{\rm{PL}}$, which is estimated from the adsorption-induced deformation measured on a different sample. The solid bulk modulus obtained from the effective stress law is $1.15$ times greater than for a solid with cylindrical pores. This can be interpreted as an indication that pores in Vycor glass are slightly ``softer'' than straight circular cylinders. Since both poroelasticity and EMT methods have sources of uncertainty, we assume that these two estimates provide a likely range of $K_{\rm{s}}$ values.  

\subsubsection{Sample from \citet{Page1995}}

The Vycor glass sample studied by \citet{Page1995} has slightly higher porosity ($\phi =0.309$) and at the same time a higher dry bulk modulus ($K_{0}=10.1$~GPa) than used by \citet{Schappert2014}, and hence considerably higher solid bulk modulus. There are no adsorption-induced deformation data for this sample, but EMT with cylindrical pores gives $K_{\rm{s}}=22.78$~GPa. Assuming that the pore geometry of the \citet{Page1995} samples is similar to that in Vycor 7930, we can multiply the EMT estimate for their sample by the factor $1.15$ to give $K_{\rm{s}}=26.2$~GPa.  

\subsection{Comparison of Gassmann Theory with Experimental Data}

\subsubsection{Constant fluid modulus}

Although the calculation of the fluid bulk modulus $K_{\rm{f}}$ is more straightforward than the solid modulus $K_{\rm{s}}$, it is a key component of the proposed method. Indeed, if we substitute the bulk modulus of bulk fluid at the pressure of the vapor phase $p$ into Gassmann equation \ref{GassmannM}, the result would be almost the same as for dry sample, in variance to the laboratory measurements. This suggests that the the value of $K_{\rm{f}}$ needs to be corrected.

Since above the capillary condensation points the fluids in the nanopores are in liquid-like form, we use the corresponding values for liquids. Hence we use $K_{\rm{f}}(0) = 1.06$~GPa for argon at $T=80$~K \citep{Gor2015compr}. Bulk argon freezes at $T=83.81$~K, and can stay liquid at $T=80$~K only when confined; therefore the properties for liquid argon are taken at saturation at higher temperatures and extrapolated to $T=80$~K \citep{Gor2014}. For n-hexane at $T = 295.75$~K at saturation we take $K_{\rm{f}} = 0.803$~GPa, following \citep{Page1995}; this value is consistent with the more recent high-quality measurements \citep{Daridon1998}. 

Two panels of Figure \ref{fig:Results} show the longitudinal moduli $M$ calculated from measurements of both wave propagation velocities and mass density as functions of relative pressure for the two considered systems. The markers show the experimental data and the black horizontal lines show the results of calculations using Eq \ref{GassmannM} with the bulk value for the fluid moduli. The resulting numbers are close to the experimental data, but the predictions do not capture the observed variation of the saturated modulus with vapor pressure $p/p_0$. 

\subsubsection{Pressure-dependent fluid modulus}

\begin{figure}[ht!]
\centerline{\includegraphics[height=2in]{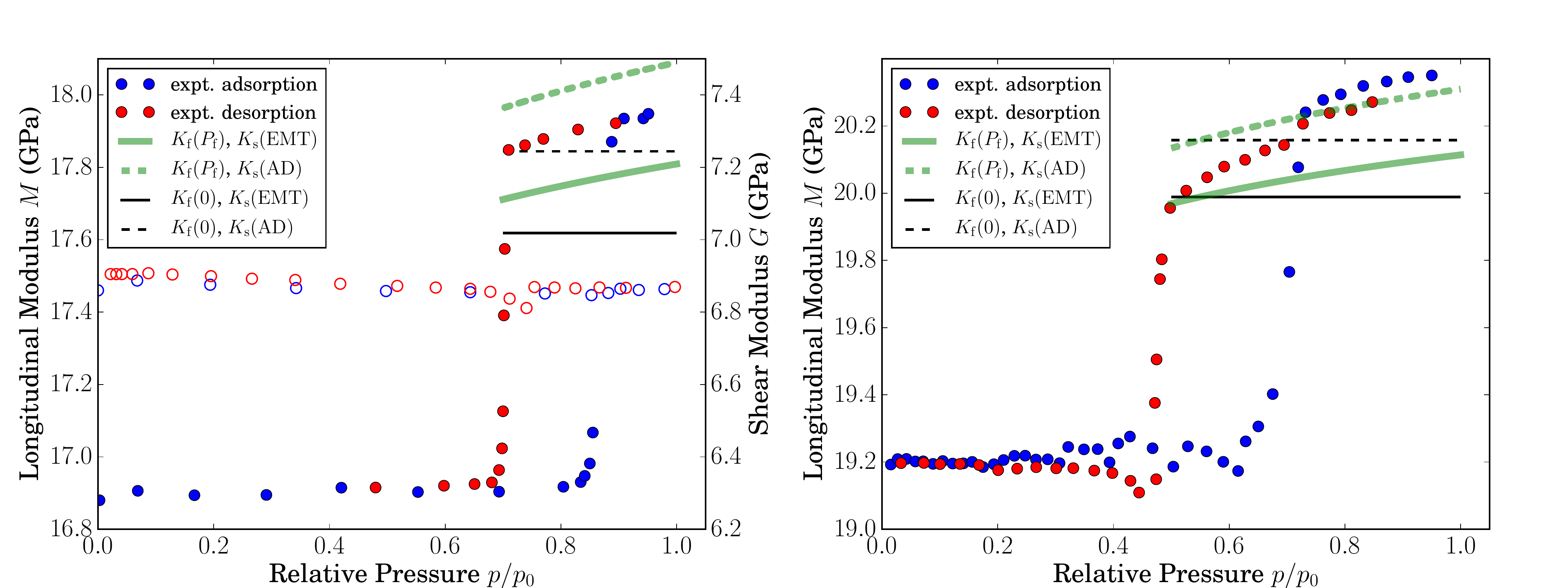}}
\caption{The change of longitudinal modulus $M$ of a sample as a function of vapor pressure during adsorption on Vycor glass. Left panel shows the data for argon adsorption at 80~K, experimental data from \citet{Schappert2014}. The right panel shows the data for n-hexane, experimental data from \citet{Page1995}. The markers show the experimental points and lines show the modulus predicted by Eq.~\ref{GassmannM}. Empty circles show shear modulus, solid circles show longitudinal modulus, blue markers correspond to the adsorption process and red ones to the desorption. The black lines show the calculations based on the constant fluid modulus, and green lines -- based on the fluid modulus corrected according to the Tait-Murnaghan equation \ref{TaitF}. Solid lines denote the results based on the $K_{\rm{s}}$ obtained from EMT and dashed lines -- based on the adsorption-induced deformation (AD) data.}
\label{fig:Results}
\end{figure}

The dependence of the saturated modulus on the vapor pressure (above the capillary condensation point) can be modeled by taking into account the effect of the solvation pressure on the bulk modulus of the confined fluid \citep{Gor2014, Gor2016Tait}. We use Tait-Murnaghan equation (Eq.~\ref{TaitF}) with the parameter $K_{\rm{f}}^{\prime} \simeq 18$ \citep{Gor2016Tait}. The calculations in \citep{Gor2016Tait} were done for isothermal modulus, therefore $K_{\rm{f}}^{\prime} = \gamma \left. K_{\rm{f}}^{\prime} \right|_{T}$, where $\gamma = 1.97$ is the heat capacities ratio \citep{Gor2014}. The Laplace pressure is readily calculated using Eq.~\ref{PLaplace} and the solid-fluid part of solvation pressure $P_{\rm{sl}}$ is the least trivial term. According to recent calculations \citep{Gor2014}, $P_{\rm{sl}} \simeq 16$~MPa for argon in 8~nm cylindrical pore at $T=80$~K. With these data we can calculate the modulus $K_{\rm{f}}$ of the fluid as a function of $p/p_0$ and substitute it into Gassmann equation \ref{GassmannM}. The results are shown with the green lines in the left panel of Figure \ref{fig:Results}.

The left panel of Figure \ref{fig:Results} also shows values of the shear modulus measured by \citet{Schappert2014}. These data show an important experimental result revealed by these authors: saturation of nanoporous glass with fluid does not appreciably change its shear modulus. Despite several observations suggesting that nanoconfinement of fluids causes an appearance of shear modulus \citep{Granick1991}, we see that for argon in 8~nm pores these effects are small, which justifies extrapolation of liquid properties to nanoconfined argon. Furthermore, this result is consistent with the Gassmann theory (Eq. \ref{GassmannG}).

Similar analysis of the n-hexane data requires the values of $K_{\rm{f}}^{\prime}$ of confined n-hexane and the solid-fluid part of the solvation pressure $P_{\rm{sl}}$ for the hexane-Vycor system. To our knowledge, none of these quantities have been calculated. Nevertheless, based on the argon simulations \citep{Gor2016Tait} showing that $K_{\rm{f}}^{\prime}$ is not affected by confinement, we can take the bulk value $K_{\rm{f}}^{\prime} \simeq 12$ calculated from the n-hexane reference data \citep{Daridon1998}. Furthermore, we can estimate the solid-fluid interaction part of the solvation pressure $P_{\rm{sl}}$ based on the solvation pressure of n-pentane adsorbed in silica pore of comparable size \citep{Gor2013}. This estimate gives $P_{\rm{sl}} \simeq 11$~MPa (see Supporting Information). Substitution of these numbers into Gassmann Eq.~\ref{GassmannM} gives the result shown as green lines in the right panel of Figure \ref{fig:Results}. Taking into account the sensitivity of the Gassmann equation to the $K_{\rm{s}}$ value, we can conclude that the resulting predictions agree with experimental data in both magnitude of the modulus and its variation with pressure. To provide an additional illustration of the sensitivity of Gassmann equation to the $K_{\rm{s}}$ (or $E_{\rm{s}}$) value, we calculated the resulting moduli for several values of this parameter from the range given by \citet{Bentz1998}. These results are plotted on Figure SI.3 in Supporting Information. 

To summarize, in order to use the Gassmann equation for nanoporous media the fluid properties should be corrected for the solvation pressure using the Tait-Murnaghan equation. The slope parameter $K_{\rm{f}}^{\prime}$ can be taken from the reference (bulk) data, which is available for any common fluid. The value of solvation pressure can be estimated from the adsorption-induced deformation experiments (see Supporting Information).

\section{Discussion}
\label{sec:Discussion}

Gassmann theory was originally derived for purely quasi-static deformations, and its use at ultrasonic frequencies may violate its key assumptions. One of these assumptions is that the average pore radius is much smaller than the viscous skin depth $\delta=\left( 2 \eta/\omega \rho_{\rm{f}} \right)^{1/2}$, where $\eta$ and $\rho_{\rm{f}}$ are dynamic viscosity and density of the pore fluid respectively, $\omega=2\pi f$ is angular and $f$ circular frequency of the propagating wave \citep{Biot:1956a,Biot:1956b,Johnson:etal:1987}. For n-hexane this condition has been checked by \citet{Page1995}. Indeed, at normal conditions, n-hexane has a density $\rho_{\rm{f}}=660$~kg$\cdot$m$^{-3}$, and dynamic viscosity $\eta=0.3 \cdot 10^{-3}$~Pa$\cdot$~s, and hence at $f=6.2$~MHz, $\delta \approx 1.5\cdot10^{-7}~\rm{m} =150$~nm, which is over an order of magnitude larger than the typical pore size in nanoporous Vycor glass. For liquid argon at 0.1~MPa and 86~K, taking $f=12$~MHz \citep{Schappert2014}, $\rho_{\rm{f}}=1.4 \cdot 10^3$~kg$\cdot$m$^{-3}$ and $\eta=0.27 \cdot 10^{-3}$~Pa$\cdot$~s \citep{Younglove1986} gives $\delta \approx 72$~nm, a smaller value, but still exceeding the pore diameter by an order of magnitude. Thus both experiments are in the low-frequency regime, which justifies the use of the Gassmann equations.

Another effect that can lead to deviations of ultrasonic measurements from the predictions of the Gassmann theory is associated with the assumption that fluid pressure is the same throughout the pore space (within a representative volume). If the pore space is interconnected, this condition is automatically satisfied at zero frequency (due to the Pascal law). However at ultrasonic frequencies this might not be the case as there may not be enough time for the pressure to equilibrate within half period of the wave. Unrelaxed pressure conditions may cause wave-induced fluid flow between pores (or areas of the pore space) of different compressibilities, resulting in significant velocity dispersion and deviation from the Gassmann's low frequency predictions \citep{Mavko:Nur:1975,Muller:Gurevich:Lebedev:2010}. However, Vycor nanoporous glass appears to be a very homogeneous material, whose pores are approximately of the same worm-like shape and similar size, see Figure~\ref{fig:porous}. This is also consistent with the very sharp jump of the longitudinal modulus at the capillary condensation point in both experiments seen in Figure~\ref{fig:Results}. Indeed, substantial variation in the pore size would lead to a gradual increase of the modulus, as capillary condensation pressure is a function of the pore diameter. This suggests that wave-induced flow effects in Vycor should be minimal and Gassmann theory should be applicable at ultrasonic frequencies. It is also known that wave-induced flow (squirt) effect associated with the presence of microcracks causes a significant deviation not only of the bulk modulus but also of the shear modulus of the saturated material from its value for the dry material \citep{Mavko1991, Gurevich2010}. Hence the fact that the shear modulus of Vycor is independent of saturation is another indirect indication that the effect of squirt in Vycor is insignificant.

In Section \ref{sec:Tait} we introduced the pressure corrections for the modulus of the fluid, while assuming that the solid modulus $K_{\rm{s}}$ does not depend on pressure. Strictly speaking, the bulk modulus of the solid constituent $K_{\rm{s}}$ changes with pressure according to the Tait-Murnaghan equation, similarly to the fluid.
\begin{equation}
\label{TaitS}
K_{\rm{s}}(P) = K_{\rm{s}}(0) + K_{\rm{s}}^{\prime} P
\end{equation}
For most of the solids the $K_{\rm{s}}^{\prime} \simeq 4-6$,  (Tables 2 and 3 in \citep{Anderson1966}), which is several times smaller than $K_{\rm{s}}^{\prime}$ for fluids. For our estimates we take the value $K_{\rm{s}}^{\prime} = 6.33$ for $\alpha$-quartz, which is relatively high. For the systems discussed above, the maximum pressure difference is $\sim 10$~MPa, for pressure $P \simeq 10$~MPa the correction term in Eq. \ref{TaitS} will be only $0.06$~GPa, and therefore could be neglected (comparing to $K_{\rm{s}}(0) =16.1$~GPa). The fact that here the solvation pressure in the pores does not cause any noticeable effect on the shear modulus of the sample (Eq. \ref{GassmannG}) also suggests that the pressure corrections for the solid moduli are insignificant.

Finally, we would like to note that although \citet{Schappert2014} reported the adsorption-induced deformation measurements for their Vycor sample, we intentionally did not use the pore-load modulus extracted from these experiments. First, \citet{Schappert2014Langmuir} themselves admit that their sample shows ``unexpected sorption-induced deformation'', which has not been observed for any other mesoporous samples \citep{Gor2017review}. Second, even if we extract the value of the pore-load modulus from these data, we get $\mathcal{M}_{\rm{PL}} = 18.9$~GPa \citep{Gor2016Bangham}. This value is smaller than $3K_0$ and hence its substitution into Eq. \ref{eq_for_K1} would give a negative value of the solid bulk modulus.

\section{Conclusion}
\label{sec:Conclusion}

We have considered the propagation of ultrasound in fluid saturated nanoporous media  within the context of classical Gassmann theory. As an example of such media we chose Vycor glass, which has well-defined channel-like pores with a narrow pore-size distribution peaking around 7-8~nm. We compared the Gassmann predictions with the two series of experimental measurements, where argon and n-hexane were used as saturating fluids. In both experiments the velocity of ultrasound propagation was measured during the vapor adsorption and the fluid adsorbed inside the pores was in liquid-like state. 

Our analysis has shown that the bulk modulus of the solid (pore walls) can be calculated from experimental data on adsorption-induced deformation or estimated using the effective medium theory, giving fairly close values. Substitution of these values and the compressibility of the bulk fluid into the Gassmann equation gives the bulk moduli of the saturated sample close to the measured values. The agreement is further improved by substituting into the Gassmann equation the fluid compressibility obtained as a function of the solvation pressure using the Tait-Murnaghan equation.

Overall, we can conclude that the saturated moduli of the nanoporous glass are consistent with the predictions of the Gassmann theory. However, at this point it would be premature to use this conclusion for practical purposes, namely for analysis of ultrasound propagation in natural nanoporous materials, e.g. coal or shale. Further validation and generalization of this conclusion requires laboratory experiments for a number of well characterized solid-fluid systems.

\acknowledgments

The data used are listed in the references, figures and Supporting Information. G.~G. thanks Drs. Klaus Schappert and Rolf Pelster for multiple detailed discussions of their experimental data. B. G. thanks the sponsors of the Curtin Reservoir Geophysics Consortium for financial support, and Dr. Stas Glubokovskikh and Yongyang Sun for useful discussions.

\newpage
\newcommand{\MPL}{\mathcal{M}_{\rm{PL}}}
\renewcommand{\thesubsection}{SI.\Roman{subsection}}

\begin{center}
\section*{Supporting Information}
\end{center}

\subsection{Solid bulk and shear moduli from the moduli of the dry porous matrix}
\label{sec:moduli}

Effective bulk and shear moduli of a solid containing randomly oriented
ellipsoidal inclusions can be computed using a non-interactive
single-scattering approximation of the effective medium theory \citep{Kuster:Toksoz:1974,Berryman1980}. For inclusions in a shape of infinite circular cylinders (sometimes called
``needles''), the corresponding equations are given in \citet{Berryman1980}, equations (23) and (24), Table 1) and reproduced more legibly in the Rock Physics Handbook \citep{Mavko:Mukerji:Dvorkin:2009}. For bulk $K_{0}$ and shear $G_{0}$ moduli of a solid with bulk modulus $K_{\rm{s}}$ and shear modulus $G_{\rm{s}}$ containing empty (dry) cylindrical inclusions of a volume fraction $\phi$, these equations read
\begin{equation}
\left( K_{\rm{s}}-K_{0}\right) \frac{K_{\rm{s}}+\frac{4}{3}G_{\rm{s}}}{K_{0}+\frac{4}{3}%
G_{\rm{s}}}=\phi K_{\rm{s}}P^{sp},  \label{KT_K}
\end{equation}%
\begin{equation}
\left( G_{\rm{s}}-G_{0}\right) \frac{G_{\rm{s}}+F_{\rm{s}}}{G_{0}+F_{\rm{s}}}=\phi G_{\rm{s}}Q^{sp},
\label{KT_G}
\end{equation}%
where 
\begin{equation}
F_{\rm{s}} =\frac{G_{\rm{s}}}{6}\frac{9K_{\rm{s}}+8G_{\rm{s}}}{K_{\rm{s}}+ 2 G_{\rm{s}}},
\end{equation}%
\begin{equation}
P^{sp}=\frac{K_{\rm{s}}+G_{\rm{s}}}{G_{\rm{s}}},
\end{equation}%
\begin{equation}
Q^{sp}=\frac{1}{5}\left( \frac{16}{3}+2\frac{G_{\rm{s}}+\gamma _{\rm{s}}}{\gamma_{\rm{s}}}%
\right) 
\end{equation}%
and%
\begin{equation}
\gamma_{\rm{s}} =G_{\rm{s}}\frac{3K_{\rm{s}}+G_{\rm{s}}}{3K_{\rm{s}}+7G_{\rm{s}}}.  \label{gamma}
\end{equation}%
Equations \ref{KT_K} - \ref{gamma} are typically used to compute
effective dry moduli $K_{0}$ and $G_{0}$ from porosity and solid moduli $%
K_{\rm{s}}$ and $G_{\rm{s}}$. Conversely, if the dry moduli $K_{0}$ and $G_{0}$ are known,
equations \ref{KT_K} - \ref{gamma} can be used to estimate solid moduli $%
K_{\rm{s}}$ and $G_{\rm{s}}$. For the sample of \citet{Schappert2014}, ($\phi
=0.28,$ $K_{0}=7.73$ GPa, $G_{0}=6.86$ GPa), eqs \ref{KT_K} - \ref{gamma} give $K_{\rm{s}}=14.13$ GPa and $G_{\rm{s}}=13.97$ GPa. For the sample of \citet{Page1995} ($\phi =0.309,$ $K_{0}=10.1$ GPa, $G_{0}=6.86$ GPa) eqs \ref{KT_K} - \ref{gamma} give $K_{\rm{s}}=22.78$ GPa and $G_{\rm{s}}=14.42$ GPa.

Strictly speaking. the non-interactive approximation is limited to low concentration of inclusions. However, numerical calculations show that for stiff inclusions like circular cylinders this approximation gives accurate results for concentrations up to and even beyond $\phi = 0.5$ \citep{Berryman1980}

\subsection{Distinction between solvation pressure and vapor pressure}
\label{sec:pressures}

\begin{figure}[ht!]
\centerline{\includegraphics[height=2in]{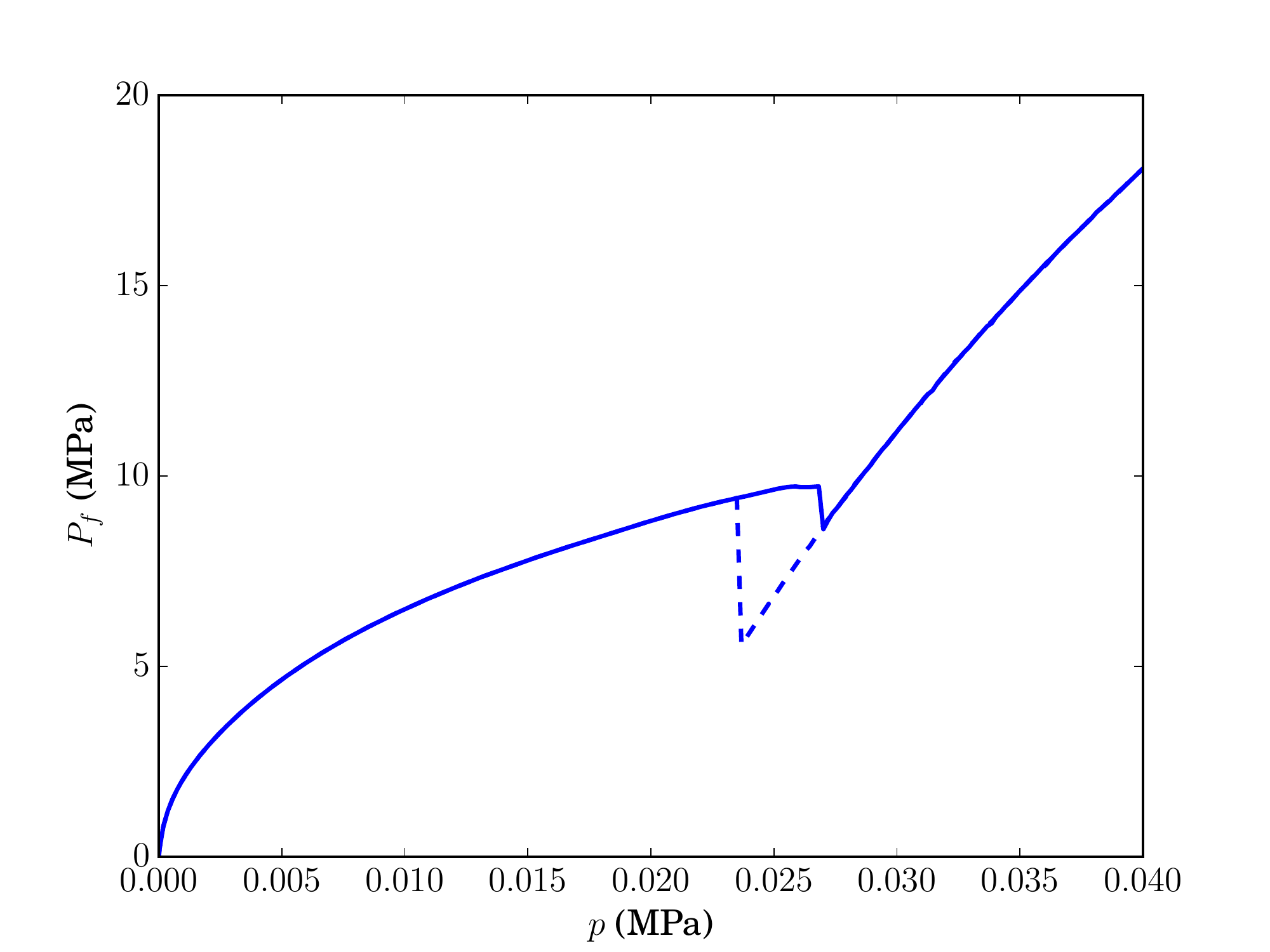}}
\caption{Solvation pressure isotherm calculated for argon adsorption in an infinite cylindrical SiO$_2$ pore at the temperature of 80~K. The calculations are done following the method from \citet{Gor2011} based on the quenched solid density functional theory (QSDFT) \citep{Ravikovitch2006}. QSDFT data is taken from \citep{Gor2014}. Solid line represents the adsorption path (increasing vapor pressure), while the dashed line shows the equilibrium desorption path (decreasing vapor pressure).}
\label{fig:argon-solvation}
\end{figure}

The solvation pressure $P_{\rm{f}}$ calculated for the argon adsorption at 80~K in cylindrical silica pore is shown in Figure~\ref{fig:argon-solvation}. The solvation pressure is shown as a function of absolute argon vapor pressure $p$, taking the saturated vapor pressure at 80~K $p_0$ as 0.037~MPa (extrapolating data to 80~K using the Antoine equation). Figure~\ref{fig:argon-solvation} shows that characteristic solvation pressure in this system exceeds the vapor pressure by 2-3 orders of magnitude. 

\subsection{Estimation of solvation pressure from the adsorption-induced deformation data}
\label{sec:solvation}

Tait-Murnaghan equation for elastic modulus of fluid requires the pressure in the fluid phase. For nano-confined fluid this pressure is referred to as ``solvation pressure'' and can be calculated theoretically, which is typically done using molecular modeling (e.g. the   case of argon \citep{Gor2014} discussed above in Section \ref{sec:pressures}). Since solvation pressure is the driving force for the adsorption-induced deformation, it can be also estimated from an experimental strain isotherm \citep{Gor2017review}. Experimental data on adsorption-induced deformation of n-hexane-Vycor system is not available, and therefore a rough estimate for the solvation pressure can be made based on the published data of n-pentane adsorption on templated mesoporous silica \citep{Prass2012,Gor2013,Balzer2015}. Although the elastic properties (pore-load modulus) of the silica materials studied in these works are significantly lower than that of Vycor, the pore sizes and surface physico-chemical properties, which determine the solvation pressure, are quite close.  

\begin{figure}[ht!]
\centerline{\includegraphics[height=2in]{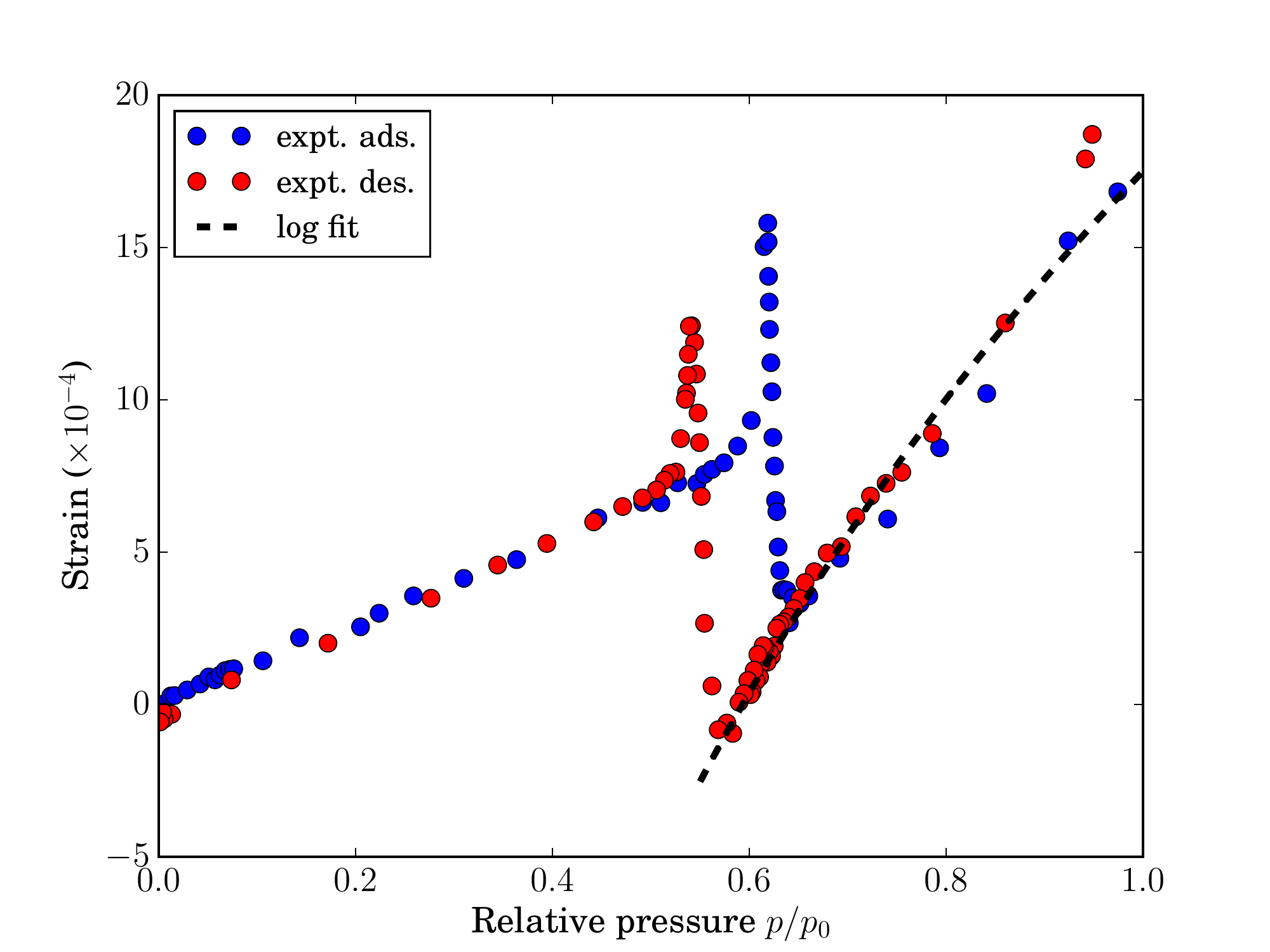}}
\caption{Linear strain of SBA-15 silica induced by n-pentane adsorption at room temperature as a function of relative vapor pressure. Markers show the experimental data and the dashed line represents the log fit of the data after the capillary condensation. Data from \citep{Gor2013}. }
\label{fig:pentane}
\end{figure}

Figure \ref{fig:pentane} gives the experimental data for adsorption-induced strain for n-pentane-SBA-15 system. SBA-15 silica has channel-like pores with a very narrow pore size distribution \citep{Zhao1998}; the mean pore size for the considered sample is 8.14 nm \citep{Gor2013}. Solvation pressure for the filled pores is given by Eqs 4 and 5 in the paper, which can be written as
\begin{equation}
\label{PsolvationSI}
P_{\rm{f}} = P_{\rm{sl}} + \frac{R_{\rm{g}} T}{V_{\rm{l}}} \log\left( \frac{p}{p_0}\right).
\end{equation}  
It is related to the experimentally-measured strain through the pore-load modulus: $\epsilon_{l} = P_{\rm{f}}/\MPL$. Fitting the part of the experimental curve with the logarithmic function provides both parameters for the system, the pore-load modulus $\MPL \simeq 6.4$~GPa and the solvation pressure for a totally filled pore (when $p=p_0$), $P_{\rm{sl}}\simeq 11$~MPa. Note that the estimate for solvation pressure of n-pentane in a different silica material with similar pore sizes (hierarchical silica-based monoliths) gives a close value of $P_{\rm{sl}}$ \citep{Balzer2015}.

\subsection{Range of the Gassmann equation predictions}
\label{sec:range}

\begin{figure}[ht!]
\centerline{\includegraphics[height=2in]{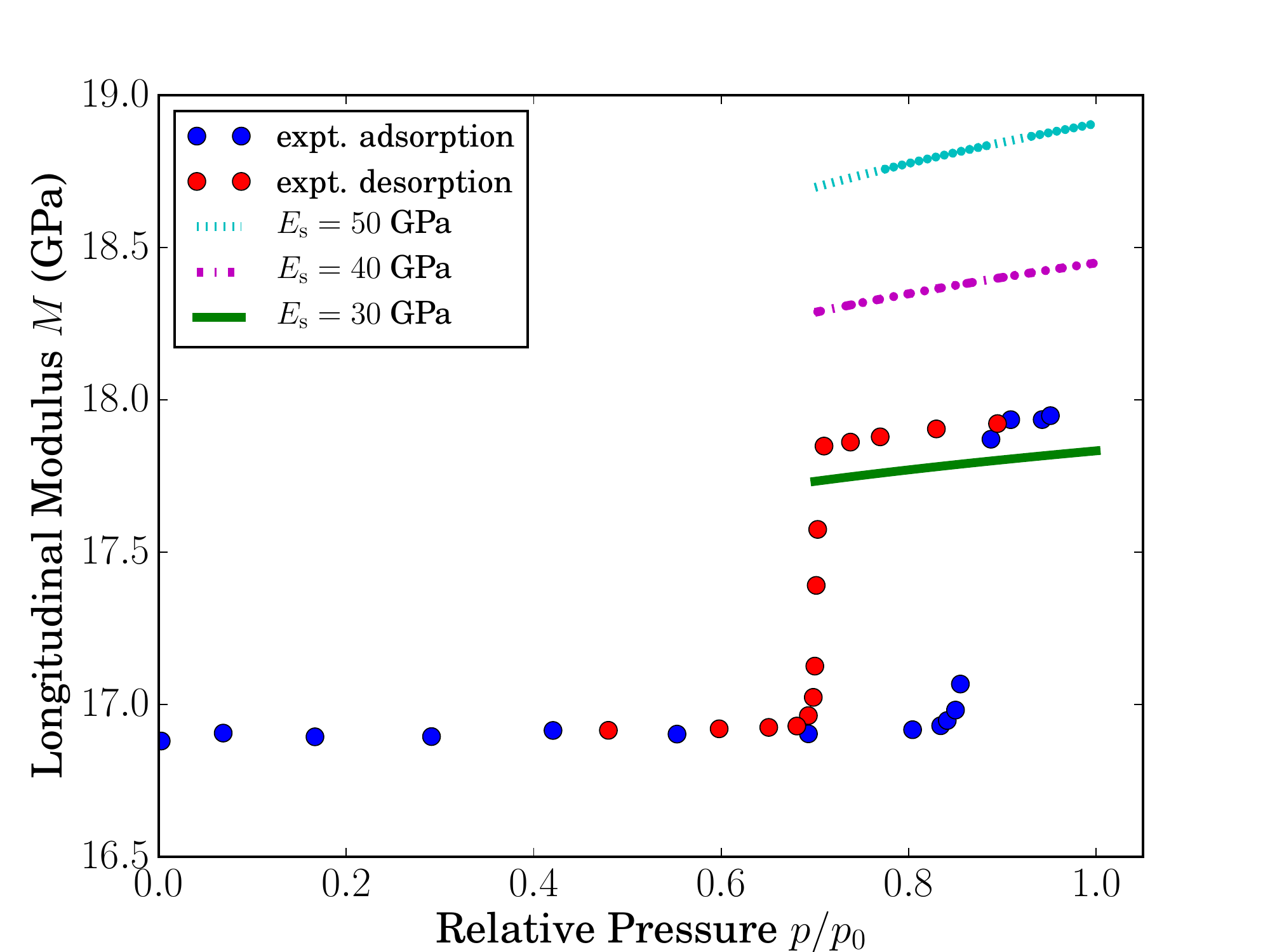} \includegraphics[height=2in]{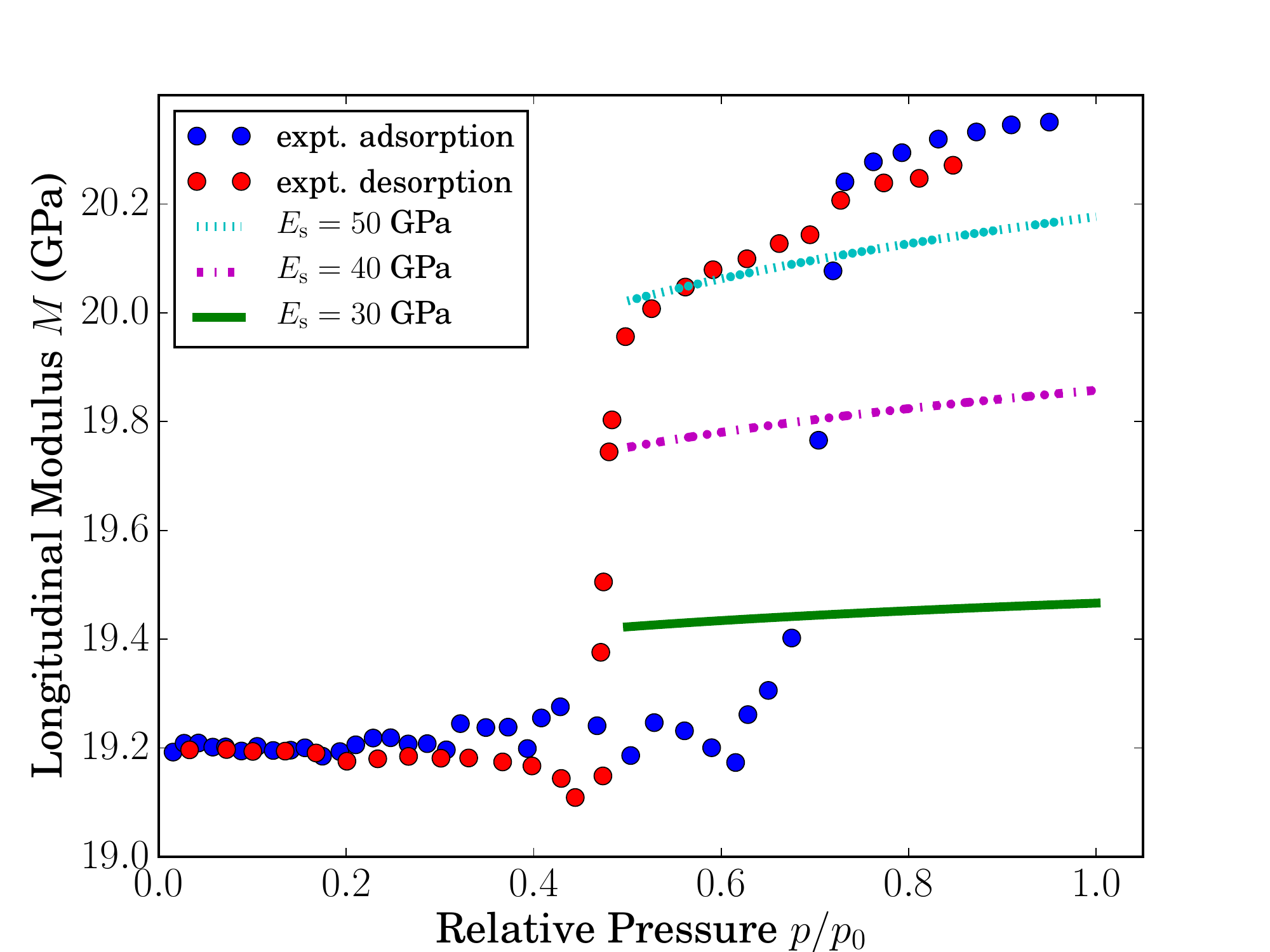}}
\caption{The change of longitudinal modulus $M$ of a sample as a function of vapor pressure during adsorption on Vycor glass. Left panel shows the data for argon adsorption at 80~K, experimental data from \citet{Schappert2014}. The right panel shows the data for n-hexane, experimental data from \citet{Page1995}. The markers show the experimental points and lines show the modulus predicted by the Gassmann equation. Blue markers correspond to the adsorption process and red ones to the desorption. Three lines in each panel correspond to three values of the Young's modulus.}
\label{fig:Bounds}
\end{figure}

Since the hydrostatic moduli of the fluids considered in the current work are noticeably smaller than the moduli of the solid, the predictions of Gassmann equation are very sensitive to the values of the moduli of the solid constitutive. The literature values for the solid properties of Vycor glass (bulk modulus $K_{\rm{s}}$ or Young's modulus $E_{\rm{s}}$) vary in a wide range \citep{Bentz1998}. In order to demonstrate the sensitivity of the results to these parameters we show the calculations for the three characteristic values of the Young's modulus $E_{\rm{s}} = 30$, $40$, and $50$~GPa, which lay in the range discussed by \citet{Bentz1998}, taking Poisson's ration $\nu = 0.15$ \cite{Scherer1986}. The results of the calculations for argon and n-hexane systems are shown in Figure~\ref{fig:Bounds}. These results provide further confirmation of the observation described in the main text that the solid constituent of the Vycor samples used in the experiments of \citet{Schappert2014} and \citet{Page1995} differ in their bulk modulus. This is consistent with the observations of \citet{Bentz1998} that different Vycor samples vary in their material properties.

\end{document}